\documentclass[apl, aps, two column,showpacs,floatfix]{revtex4-1}
\usepackage{comment}
\usepackage{xcolor}
\usepackage{multirow}
\usepackage{graphics,graphicx,epsfig}
\usepackage[caption=false]{subfig}
\usepackage{amssymb,amsmath}
\usepackage{epstopdf}
\usepackage{amsmath}
\usepackage{bbold}
\usepackage{mathtools}
\usepackage{hyperref}
\hypersetup{colorlinks=true, citecolor = blue, linkcolor=blue, filecolor=magenta, urlcolor=blue}
\begin{document}
\title{Quantum synchronization between two spin chains using pseudo-bosonic equivalence}

\author{Jatin Ghildiyal*, Manju, Shubhrangshu Dasgupta, Asoka Biswas}

\affiliation{ Department of Physics, Indian Institute of Technology Ropar, Rupnagar, Punjab 140001, India}

\begin{abstract}
Quantum synchronization among many spins is an intriguing domain of research. In this paper, we explore the quantum synchronization of two finite chains of spin-1/2 particles, via a nonlinear interaction mediated by a a central intermediary spin chain. We introduce a novel approach using the Holstein-Primakoff transformation to treat the spin chains as pseudo-bosonic systems and thereby applying the synchronization criteria for harmonic oscillators. Our theoretical framework and numerical simulations reveal that under optimal conditions, the spin chains can achieve both classical and perfect quantum synchronization. We show that quantum synchronization is robust against variations in the number of spins and inter-spin coupling, though may be affected by thermal noise. This work advances the understanding of synchronization in multi-spin systems and introduces a generalized synchronization measure for both bosons and fermions.

\end{abstract}
\maketitle
\section{INTRODUCTION}
Recent advances in quantum synchronization, at the crossroads of quantum physics and nonlinear dynamics, offer an exciting domain of research \cite{pikovsky2001universal,strogatz2018nonlinear, huygens1897oeuvres}. Usually, in the classical regime, one explores synchronized patterns in coupled oscillators \cite{huygens1897oeuvres}, which persists even at different natural frequencies of these oscillators \cite{pikovsky2001universal,strogatz2018nonlinear, huygens1897oeuvres,arenas2008synchronization}. In the quantum regime, on the other hand, these studies can be extended to a few-level systems, leading to a more generalized approach to synchronization. Quantum synchronization has been studied in various platforms, e.g., van der Pol oscillators, Josephson junction arrays, spin torque nano-oscillators, and optomechanical systems \cite{heinrich2011collective,holmes2012synchronization,zhang2012synchronization,PhysRevLett.111.103605, wiesenfeld1996synchronization,kaka2005mutual, shim2007synchronized,shim2007synchronized,manzano2013avoiding,manzano2013synchronization,giorgi2012quantum}. 

These studies hold significant promise for applications in quantum information processing, communication, and control. An interesting relation between quantum Fisher information and quantum synchronization has recently been outlined in \cite{vaidya2024quantumsynchronizationdissipativequantum}, which makes synchronization relevant, even in the domain of quantum sensing.

While quantum synchronization is traditionally studied among two or more coupled oscillators, spin synchronization is comparably a newer concept. In this context, we note that quantum synchronization between two spatially separated spin-1/2 qubit clocks has been proposed in \cite{PhysRevLett.85.2006,PhysRevLett.85.2010,shi2022clock}, which refers to adjusting their time-difference by taking advantage of entanglement shared between them. These ideas have been demonstrated using photons \cite{liu2021quantum,refId0} and nuclear magnetic resonance \cite{PhysRevA.70.062322}. Multiparty clock synchronization has also been proposed \cite{PhysRevA.66.024305,PhysRevA.84.014301,PhysRevA.86.014301} and experimentally demonstrated \cite{kong2018demonstration}. The effect of decoherence on such synchronization has been studied in \cite{noorbakhsh2024quantum}. Clock synchronization without entanglement is also proposed in \cite{PhysRevA.72.042301}.

It is however shown that to synchronize a quantum system to an external driving, the relevant Hilbert space should have a minimum size of 3  \cite{PhysRevLett.121.053601}, as a single qubit cannot be entrained due to the lack of a limit cycle. There have been several reports of synchronizing a single qubit, e.g., when coupled to a driven dissipating oscillator \cite{PhysRevLett.100.014101}  or via a mechanical resonator \cite{PhysRevA.107.013528}, or in a trapped ion \cite{PhysRevResearch.5.033209}. That a single qubit can be understood as containing a valid limit cycle has been discussed in \cite{PhysRevA.101.062104}. Synchronization of a single spin has further been demonstrated using Nitrogen-vacancy spin qubit in a radio-frequency field \cite{PhysRevLett.112.010502} and emulated in IBM Q system \cite{PhysRevResearch.2.023026}.

 Quantum synchronization between two spins, on the other hand, is interpreted in terms of constant phase relations between the corresponding limit cycles or the relative phase of their respective Bloch vectors \cite{PhysRevA.94.032336}. It is shown that two qubits can be synchronized via their coupling to a common bath \cite{PhysRevA.88.042115}, or via a coherent coupling with each other, while each being coupled to a separate bath \cite{PhysRevA.109.033718} or due to collision \cite{PhysRevA.100.012133}, without any external driving or when coupled to a driven dissipative resonator \cite{PhysRevB.80.014519}. Phase synchronization of two nuclear spins has been experimentally verified in terms of Husimi Q-function \cite{PhysRevA.105.062206}. 
 
 Two-spin synchronization has also been described as a persistent oscillation of the eigenmodes of the corresponding Liouvillean in the presence of decay. This idea was demonstrated in a system of three spins, the Hamiltonian of which conserves total magnetization \cite{10.21468/SciPostPhys.12.3.097}. Two spins of this trio exhibit antisynchronization in the transverse components of the local spins \cite{buvca2022algebraic}, thanks to the inherent dynamical and permutation symmetry. Note that such synchronization between two qubits has been explored via their common coupling with a third qubit. The mediated synchronization between two qubits has also been explored in an ion trap in the presence of a damped normal mode of the collective vibration of the ions \cite{PhysRevA.95.033423}.

While there have been several studies of quantum synchronization of a single qubit and two qubits, there has been very little investigation on many-qubit synchronization. 
Li et al. investigated synchronization in a few-spins system with non-local dissipation, revealing stable oscillatory behaviors in long-time dynamics without external driving \cite{PhysRevA.107.032219}. Stable (anti)synchronization between local spin observables can be induced by noise, under specific conditions in an isolated quantum many-body system \cite{goldobin2005synchronization,schmolke2022noise}. A generalized concept of synchronization, namely, measure synchronization has been introduced \cite{PhysRevA.90.033603} to study the coordinated dynamics of two many-body systems, coupled via contact particle-particle interactions. Correlated phase dynamics of two mesoscopic ensembles of atoms also exhibit synchronization, via their common coupling to an optical cavity \cite{PhysRevLett.113.154101}. 

In this paper, we demonstrate how two long chains of spin-1/2 particles can be quantum synchronized. A spin chain is essentially a one-dimensional (1D) lattice, with distance-dependent exchange interactions between any two spins. We consider only the nearest neighborhood interactions, as in usual 1D Ising models. Such models have been previously studied for entanglement propagation and quantum phase transitions \cite{pappalardi2018scrambling,abanin2019colloquium,ramos2020optical}. In our case, 
these two chains interact nonlinearly via their common coupling to another intermediary spin-chain. In the previous works, spin synchronization has been characterized in terms of either Husimi Q-function \cite{PhysRevA.105.062206,PhysRevLett.121.053601,PhysRevLett.121.063601,PhysRevA.101.062104,PhysRevResearch.5.033209,PhysRevA.109.033718} or Pearson spin-spin correlation function \cite{PhysRevA.100.012133}. We employ a different approach - we treat the spin chain as a pseudo-bosonic system. We introduce an equivalent pseudo-bosonic operator for the collective spin of the finite-size chain using the Holstein-Primakoff transformation \cite{PhysRev.58.1098}. This enables us to use the standardized criterion of quantum synchronization of two harmonic oscillators \cite{PhysRevLett.111.103605}. Note that the transformation `from-spin-to-boson' explicitly reveals the inherent nonlinearity in the system which leads to the synchronization. It is anticipated that quantum synchronization is a certain manifestation of quantum correlations. In fact the criterion used in \cite{PhysRevLett.111.103605} originates from the Heisenberg uncertainty principle, in the context of EPR-like variables \cite{garg2023quantum,dasgupta2023entanglement}.

The paper proceeds as follows. In Section II, we introduce the model, including its mean field approximation and necessary dynamical equations. We discuss the main results of synchronization in terms of the limit cycles and synchronization markers, in Section III. We conclude the paper in Section IV. 

\section{MODEL AND EQUATION OF MOTION}
 We will study synchronization between two distinct finite spin chains, consisting of $N_1$ and $N_2$ spin-1/2 particles [each depicted as red dots in Fig. 1(a)], respectively. These chains are weakly coupled to a common spin chain [the spins of which are depicted as black dots in Fig. 1(a)], via the site-specific spin-spin coupling (coupling constant $g_j$, $j\in 1,2$). This central chain acts as an intermediary to build up synchronization between two other chains. We further assume a nearest-neighbour coupling  between the spins, with a coupling constant $\lambda^2 \omega_j$ ($j\in 1,2$), where $\omega_j$ is the frequency of each spin in the $j$th chain and $0
\le \lambda \le 1$ is a weight factor. Here $\lambda \sim 0$ ($\lambda \lesssim 1$) refers to weak (strong) coupling among the spins. 

For an odd number of spins, the intermediary chain exhibits a ground state doublet. At low temperatures, this chain is assumed to remain confined to this doublet and hence may be considered as an equivalent single spin [depicted by a black circle in Fig. 1(b)]. The coupling between two chains is now essentially mediated via this single spin, with a coupling constant modified by a factor of $1/\sqrt{N_j}$. The equivalent Hamiltonian is then expressed as follows:

\begin{equation}
H = H_{S}+\sum_{j=1}^2(H_{SB_j}+H_{B_j})\;,
\end{equation}
where
\begin{small}
\begin{eqnarray}
H_S &=& \frac{\hbar \omega_0}{2} \sigma_{z}^0 \;, \nonumber \\
H_{SB_j} &=& \frac{\hbar g_j}{2 \sqrt{N_j}} \sum_{i=1}^{N_j} \left( \sigma_{x}^0 \sigma_{j x}^i + \sigma_{y}^0 \sigma_{j y}^i + \sigma_{z}^0 \sigma_{j z}^i \right) \;, \nonumber \\
H_{B_j} &=& \frac{\hbar \omega_j}{2 N_j} \sum_{i=1}^{N_j} \left\{ \frac{1}{2} \sum_{\substack{k=1 \\ k \neq i}}^{N_j} \left[ \lambda^2 (\sigma_{j x}^i \sigma_{j x}^k + \sigma_{j y}^i \sigma_{j y}^k) \right] + \sigma_{j z}^i \right\} \;.
\end{eqnarray}
\end{small}

\begin{figure}[ht]
\includegraphics[height=11.2cm,width=20.3cm]{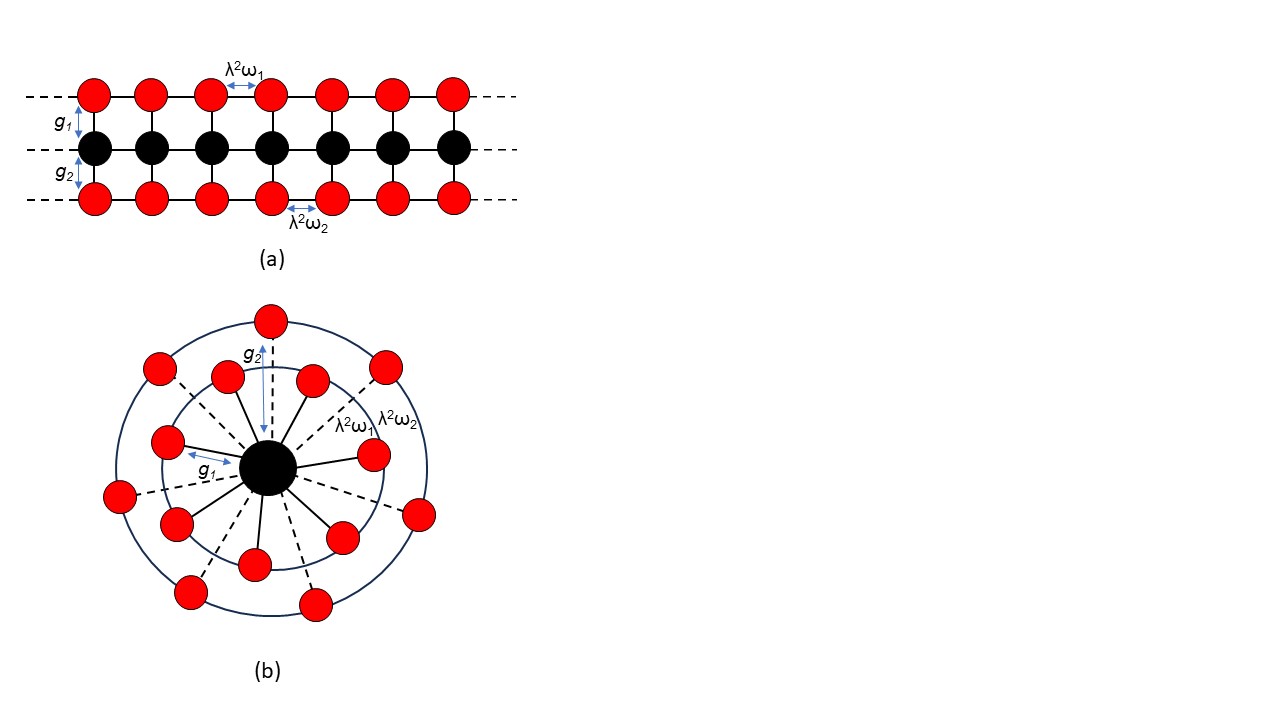}
\caption{Schematic of a generic (a) Spin Chain Model, (b) Central-Spin Model}
\label{fig}
\end{figure}

Next, we invoke the collective spin angular momentum operator, \( J_{jk'} = \frac{1}{2} \sum_{i=1}^{N_j} \sigma_{jk'}^i \) (where \( k' \in \{\text{x}, \text{y}, \text{z}, +, - \}\) and \( j \in \{1, 2\} \)). Then the different parts of $H$ can be rewritten as 
\begin{small}
\begin{eqnarray}
H_{SB_j} & =&\frac{\hbar g_j}{\sqrt{N_j}}\left(\sigma_{x}^0 J_{j x}+\sigma_{y}^0 J_{j y}+\sigma_{z}^0 J_{j z}\right)\;, \nonumber\\
H_{B_j} & =&\hbar \omega_j\left[\lambda^2\left(\frac{J_{j+} J_{j-}}{N_j}-\frac{\mathbb{1}}{2}\right)+(1-\lambda^2)\frac{J_{jz}}{N_j}\right]\;
\end{eqnarray}
\end{small}
Note that the Hamiltonian $H_{SB_j}$ can be rewritten as $\frac{\hbar g_j}{\sqrt{N_j}} \vec{\sigma^0}.\vec{J_j}$, which represents an effective coupling between the central spin and the collective spin. For brevity, we will replace $\sigma^0$ with $\sigma$. 

As we mentioned in the Introduction, we will analyze spin synchronization in terms of bosons. In this regard, we introduce bosonic operators using the Holstein-Primakoff (HP) transformation \cite{PhysRev.58.1098, PhysRevA.96.052125,PhysRevA.106.032435}:
\begin{small}
\begin{eqnarray}
    J_{j+} &=& \sqrt{N_j} b_j^\dagger \left(1 - \frac{b_j^{\dagger} b_j}{2N_j}\right)^{1/2}\;,\nonumber\\
J_{j-} &=& J_{j+}^\dag=\sqrt{N_j} \left(1 - \frac{b_j^{\dagger} b_j}{2N_j}\right)^{1/2} b_j\;,\nonumber\\ 
J_{jz}& =& b_j^{\dagger} b_j - N_j\;.
\end{eqnarray} 
\end{small}
Then the Hamiltonian gets modified to ($\hbar = 1$)

\begin{small}
\begin{eqnarray}
H_{SB_j}&=&g_j\left[\left\{\frac{\sigma_{+}}{2}\left(1-\frac{b_j^{\dagger} b_j}{4 N_j}\right) b_j + {\rm h.c}\right\}+ \frac{\sigma_z}{\sqrt{N_j}}\left(b_j ^{\dagger} b_j- N_j\right)  \right]\;, \nonumber\\
H_{B_j} &= & {\omega_j}\left[b_j^{\dagger}b_j B_j-\lambda^2\frac{(b_j^{\dagger}b_j)^2}{2N_j}+\left(\frac{\lambda^2}{2}-1\right)\mathbb{1}\right]\;, \label{eq5}
\end{eqnarray}
\end{small}
where  \(B_j = \lambda^2 \left(1 - \frac{1}{2N_j}\right) + \frac{1}{N_j} \) for \( j \in \{1, 2\} \). Here, $b_j$ and $b_j^{\dagger}$ are the bosonic annihilation and creation operators with the property $\left[b_j, b_j^{\dagger}\right]=1$. In the Hamiltonian $H_{SB_j}$ in Eq. (\ref{eq5}), we have considered the terms up to the order of $1/N_j$, while expanding the series of $(1-b_j^\dag b_j/2N_j)^{1/2}$. Note that for an infinite spin chain $(N_j\rightarrow \infty) $, the above HP transformation makes a collective spin annihilation operator equivalent to a bosonic annihilation operator. In this paper, we are however focusing on finite spin chains. 

Incorporating the interaction picture and using the Baker–Campbell–Hausdorff formalism, the refined Hamiltonian $H^{\prime}(t)$ can be expressed as $(\iota = \sqrt{-1})$
\begin{equation}
\begin{aligned}
H^{\prime}(t) &= \sum_{j=1}^2 \left\{g_j \exp \left[\iota t\left(\omega_0 - \omega_j B_j\right)\right] \sigma_{+} \left(1 - \frac{b_j^{\dagger} b_j}{4 N_j}\right)\right. \\
& \left.\quad \times \exp \left[\iota \lambda^2 \omega_j t\left(\frac{b_j^{\dagger} b_j + b_j b_j^{\dagger}}{2N_j}\right)\right] b_j+ \text{h.c.}\right\} \\
& \quad + g_j \frac{\sigma_z}{\sqrt{N_j}} \left(b_j^{\dagger} b_j - N_j\right) 
\end{aligned}
\end{equation}

We assume that the central spin mediates an indirect interaction between two spin chains. The transition probability between the two levels of the central spin is assumed to remain negligible at the time scale of the evolution of the spin chain operators $b_j$s, leading to the adiabatic elimination of the central spin operators. Consequently, the time derivatives of the raising and lowering operators $\dot{\sigma}_\pm$ are both zero. 
We thereby obtain the expression of $\sigma_\pm$, in which we replace $\sigma_z$ with its average value $\langle \sigma_z\rangle$. The Hamiltonian $H^{\prime}(t)$ thus takes the following form:
\begin{equation}
\begin{small}
\begin{aligned}
H^{\prime}(t) =\sum_{j=1}^2 \frac{\langle\sigma_z\rangle}{X} g_j^2 b_j^{\dagger}\left(1-\frac{b_j^{\dagger} b_j}{4N_j}\right)^2 b_j + ({P_1}{P_2^\dagger}+ {\rm h.c.})+\langle\sigma_z\rangle\;,
\end{aligned}
\end{small}
\end{equation}
where
\begin{small}
\begin{eqnarray}
P_j &=& g_j b_j^{\dagger}\exp \left[ R_j\left(b_j^{\dagger}b_j + b_j b_j^{\dagger}\right) \right] \left(1 - \frac{b_j^{\dagger} b_j}{4N_j} \right) \exp\left(-\iota \omega_j B_j t\right)\;, \nonumber\\
R_j &=& \iota \omega_j \left(\frac{\lambda^2}{2N_j}\right) t\;, \nonumber\\
X &=& \sum_{j=1}^2 g_j\frac{1}{\sqrt{N_j}}\left(\langle b_j^{\dagger} b_j \rangle - N_j \right) \;.
\end{eqnarray}
\end{small}
Note that the above Hamiltonian is inherently nonlinear. Here we assumed $\langle \sigma_z\rangle$ to be constant.  

We study the spin-chain synchronization in terms of that of the bosons $b_j$. This is the key difference between our approach and other relevant works. We first obtain Langevin's equations for these annihilation operators, as presented in Appendix A. To obtain these equations (\ref{b1eqn}) and (\ref{b2eqn}), we used the transformations \( b_1 \to \widetilde{b_1} \exp(-\iota \theta t) \) and \( b_2 \to \widetilde{b_2} \exp(\iota \theta t) \) where \( \theta = \left[\left(\frac{\omega_2}{2}\right)B_2 - \left(\frac{\omega_1}{2}\right)B_1\right] \).
For the sake of brevity, we will denote \(\widetilde{b_1}\) and \(\widetilde{b_2}\) by \( b_1 \) and \( b_2 \), respectively in the later sections.
Further, $\gamma_l$ and $\gamma_{nl}$ are the linear and nonlinear dissipation rates of the system, respectively. Accordingly, the input noise operators ${b}_{in}^{(1)}$ and ${b}_{in}^{(2)}$ exhibit the following two-time correlation functions:

\begin{equation}
\begin{small}
\begin{aligned}
\left\langle \hat{b}_{\text{in}}^{(j)}(t) \hat{b}_{\text{in}}^{(j)\dagger}(t') \right\rangle &=  \left(\bar{n}_{m} + 1 \right) \delta(t - t') \\
\left\langle \hat{b}_{\text{in}}^{(j)\dagger}(t) \hat{b}_{\text{in}}^{(j)}(t') \right\rangle &=  \bar{n}_{m} \delta(t - t')\;.
\end{aligned}
\end{small}
\end{equation}
Here $\bar{n}_m$ is the average phonon number of the thermal bath, common to both the spin chains.

\subsection{Solution in mean-field approximation}
Due to the analytical complexity of solving the equations (\ref{b1eqn}) and (\ref{b2eqn}), we employ the mean-field approximation to simplify calculations. In the limit of large excitation of the bosonic modes, the relevant operators can be expressed as the sum of their mean values and quantum fluctuations near the mean values, i.e.,
${b}_j \rightarrow \left\langle{b}_j\right\rangle + \delta{b}_j = {\beta}_j + \delta{b}_j$. The equations for these mean values $\beta_j$ are written in Appendix B. The Eqs. (\ref{beta1eqn}) and (\ref{beta2eqn}) would represent the independent damped oscillation of the two modes, if $\langle \sigma_z\rangle = 0$, i.e., if the central spin is prepared in an equal superposition of its bare states, namely, $|0\rangle$ and $|1\rangle$. This would be the only possibility as the condition $\langle \sigma_z\rangle = 0$ cannot be achieved in thermal equilibrium. Note that the states $|0\rangle$ and $|1\rangle$ are themselves many-spin states.

\begin{figure*}

     \subfloat[\label{pqa}]{%
		\includegraphics[height=4.7cm,width=0.33\linewidth]{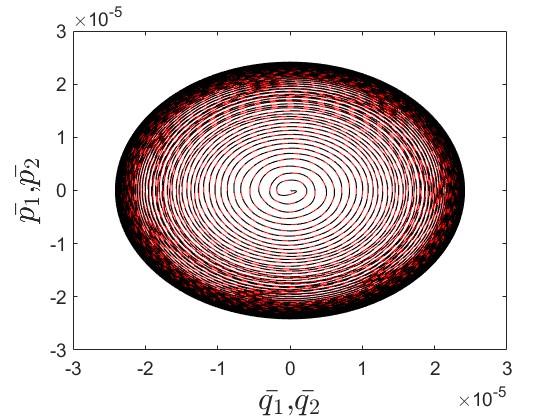}%
	 }
      \subfloat[\label{q1q2a}]{%
		\includegraphics[height=4.7cm,width=0.33\linewidth]{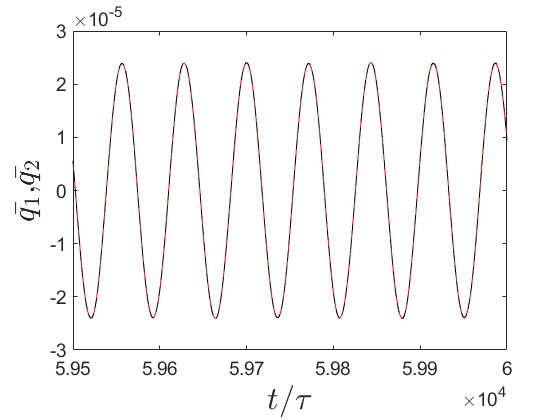}%
	 }
        \subfloat[\label{p1p2a}]{%
		\includegraphics[height=4.7cm,width=0.33\linewidth]{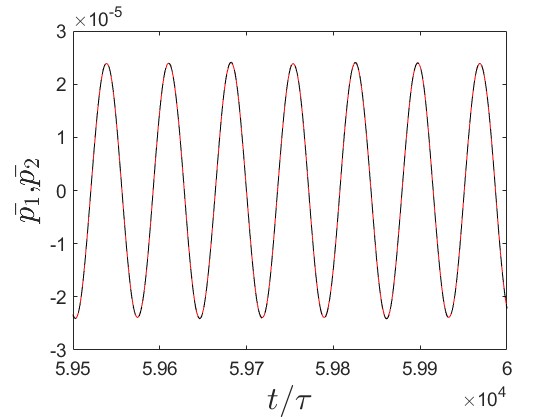}
	}\\
        \subfloat[\label{pqb}]{%
	    \includegraphics[height=4.7cm,width=0.33\linewidth]{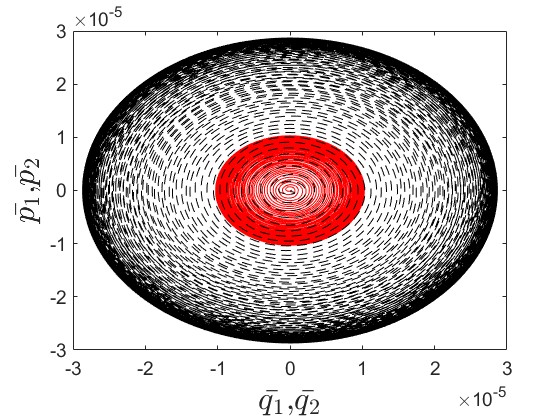}%
        }
        \subfloat[\label{q1q2b}]{%
        \includegraphics[height=4.7cm,width=0.33\linewidth]{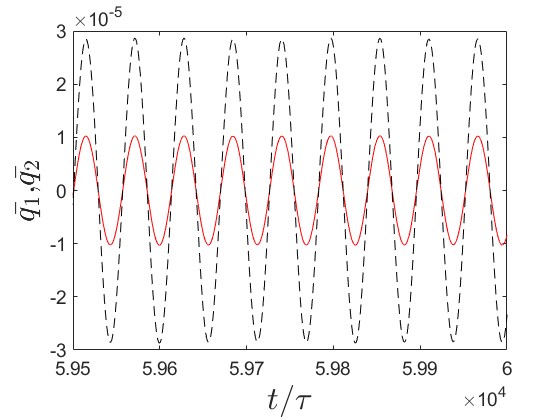}%
        }
      \subfloat[\label{p1p2b}]{%
		\includegraphics[height=4.7cm,width=0.33\linewidth]{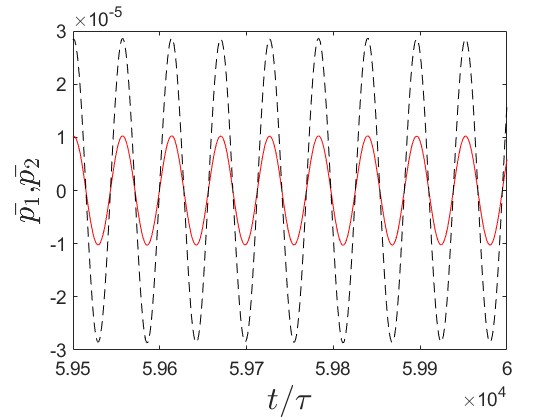}
	  }\\

     \subfloat[\label{pqc}]{%
		\includegraphics[height=4.7cm,width=0.33\linewidth]{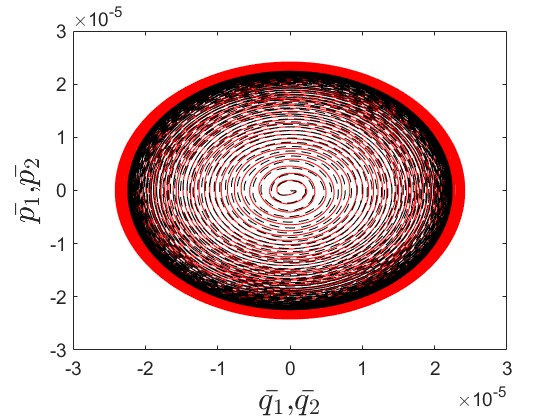}%
	}
	  \subfloat[\label{q1q2c}]{%
		\includegraphics[height=4.7cm,width=0.33\linewidth]{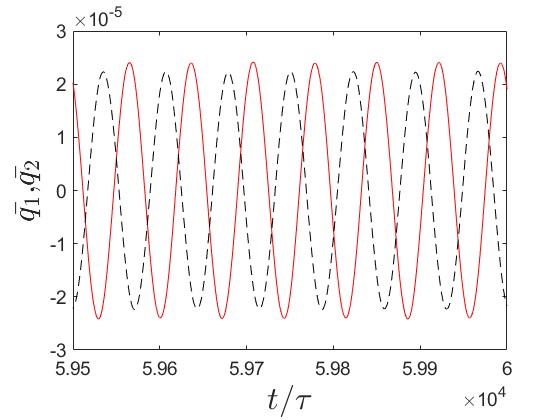}%
	}
	  \subfloat[\label{p1p2c}]{%
	    \includegraphics[height=4.7cm,width=0.33\linewidth]{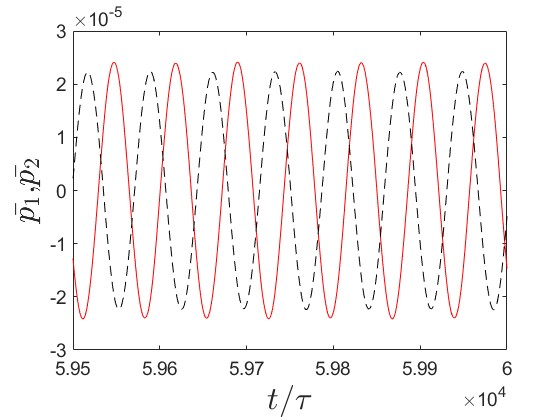}%
        }\\

 
\caption{(a) Limit-cycle trajectories in the $\Bar{q_{1}}\leftrightharpoons \Bar{p_{1}}$ (red) and $\Bar{q_{2}}\leftrightharpoons \Bar{p_{2}}$ (black) spaces, (b) Variation of the mean values $\Bar{q_{1}}$ (red) and $\Bar{q_{2}}$ (black), (c) the mean values $\Bar{p_{1}}$  (red) and $\Bar{p_{2}}$ (black). The parameters chosen are $\lambda=0$, $N_1=N_2=5$.(d) Limit-cycle trajectories in the $\Bar{q_{1}}\leftrightharpoons \Bar{p_{1}}$ (red) and $\Bar{q_{2}}\leftrightharpoons \Bar{p_{2}}$ (black) spaces, (e) Variation of the mean values $\Bar{q_{1}}$ (red) and $\Bar{q_{2}}$ (black), (f) the mean values $\Bar{p_{1}}$  (red) and $\Bar{p_{2}}$ (black). The parameters chosen are $\lambda=0$,  $N_1=10$, $N_2=5$. (g) Limit-cycle trajectories in the $\Bar{q_{1}}\leftrightharpoons \Bar{p_{1}}$ (red) and $\Bar{q_{2}}\leftrightharpoons \Bar{p_{2}}$ (black) spaces, (h) Variation of the mean values $\Bar{q_{1}}$ (red) and $\Bar{q_{2}}$ (black), (i) the mean values $\Bar{p_{1}}$  (red) and $\Bar{p_{2}}$ (black). The parameters chosen are $\lambda=0.2$, $N_1=N_2=5$ and $\phi=  1.049$. All the other parameters are
the same in all the above figures: $g_1=1.5$, $g_2=2.4$, $\langle\sigma_z\rangle=-0.1$, $ \omega_{1}=1$, $\omega_{2}=0.8$,$\gamma_l=0.001$,$\gamma_{nl}=0.002$.} 
\label{limitcycle}
\end{figure*}

To calculate the desired marker $S_q$ for quantum synchronization, we need to solve for the quadrature fluctuations of the oscillators. Solving equations (\ref{b1eqn}) and (\ref{b2eqn}) becomes more convenient by replacing relevant operators and input noise operators with their quadratures: $\delta q_j=\frac{1}{\sqrt{2}}\left(\delta b^{\dagger}_j+\delta b_j\right)$, $\delta p_j=\frac{\iota}{\sqrt{2}}\left(\delta b^{\dagger}_j-\delta b_j\right)$, $ q_{in}=\frac{1}{\sqrt{2}}\left( b_{in}^{\dagger}+ b_{in}\right)$, and $ p_{in}=\frac{\iota}{\sqrt{2}}\left( b_{in}^{\dagger}- b_{in}\right)$. In the limit of negligible higher order fluctuations, we obtain the Eqs. (\ref{c1}) which are linearized equations for the quantum fluctuations. We have included the exact forms of these 
 equations in Appendix C.

\begin{figure*}

       \subfloat[\label{Sq}]{%
		\includegraphics[height=4.7cm,width=0.33\linewidth]{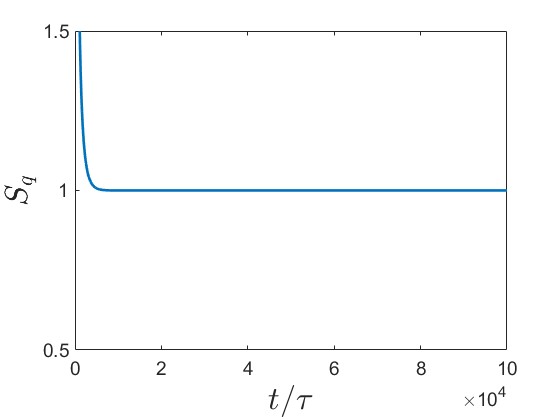}
	}
       \subfloat[\label{synchronization temp}]{%
		\includegraphics[height=4.7cm,width=0.33\linewidth]{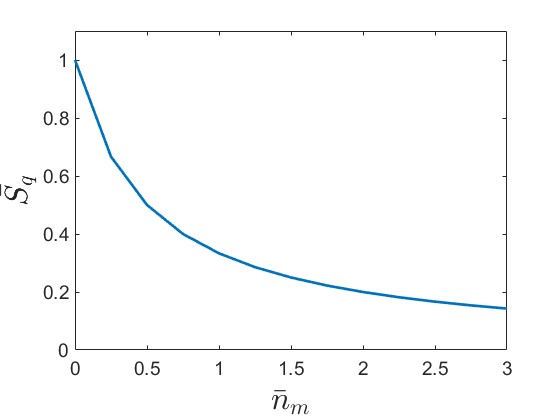 }
	}

\caption{Variation of $S_{q}$, with respect to (a) time $t$ (in the units of $\tau=1/\omega_{1}$) and  Variation of $\bar{S}_{q}$,with respect to (b)  the mean phonon number $\bar{n}_{m}$ of the environment. The other parameters are
 $g_1=1.5$, $g_2=2.4$ , $\langle\sigma_z\rangle=-0.1$, $ \omega_{1}=1$, $\omega_{2}=0.8$, $\gamma_l=0.001$, $\gamma_{nl}=0.002$, $\phi=  1.049$.}
\end{figure*}

Therefore, the fluctuation equations take a simpler form as given by 
\begin{equation}
 \dot{Y}(t) = M Y(t) + N(t) + F(t)   \;,
\end{equation}
where $Y(t)^\intercal=(\delta q_1,\delta p_1, \delta q_2, \delta p_2)$ and \begin{equation}
    M = [M_{ij}]_{4\times4}
\end{equation}
is a time-dependent coefficient matrix. Here, we neglect the  term $F(t)$ (as given in Appendix C) in the later part of this paper. The vector $N(t)$ containing the noise terms is given below:

\begin{equation}
N(t)^{T}=( \alpha_1, \alpha_2, \alpha_3, \alpha_4)\;,
\end{equation}

where 

\begin{align*}
\alpha_1 &=  {\rm Re}(U_1) q_{in}^{(1)}-{\rm Im}(U_1) p_{in}^{(1)}, \\ 
\alpha_2 &=  {\rm Im}(U_1) q_{in}^{(1)}+{\rm Re}(U_1) p_{in}^{(1)}, \\ 
\alpha_3 &= {\rm Re}(U_2) q_{in}^{(2)}-{\rm Im}(U_2) p_{in}^{(2)}, \\
\alpha_4 &= {\rm Im}(U_2) q_{in}^{(2)}+{\rm Re}(U_2) p_{in}^{(2)},
\end{align*}

and 
\begin{align*}
U_j &= \left[\sqrt{\gamma_l} + \frac{1}{N_j} \sqrt{\gamma_{nl}} \left(2|\beta_j|^2 - \beta_j^2\right)\right], \quad (j \in \{1, 2\}).
\end{align*}

To ascertain the degree of quantum synchronization exhibited between mechanical oscillators, we adopt a figure of merit originally proposed by Mari et al. \cite{PhysRevLett.111.103605}:
\begin{equation}
S_{c}(t) = \left\langle q_{-}^{2}(t) + p_{-}^{2}(t) \right\rangle^{-1} \label{eq:sc}
\end{equation}
where $q_{-}(t)$ and $p_{-}(t)$ denote the synchronization errors defined as:
\begin{eqnarray}
q_{-}(t)& =& \frac{1}{\sqrt{2}}\left[ q_{1}(t) - q_{2}(t) \right]\nonumber\\
p_{-}(t) &=& \frac{1}{\sqrt{2}}\left[ p_{1}(t) - p_{2}(t) \right]\;.
\end{eqnarray}

To identify the fluctuations, the above variables are  redefined with respect to their mean values as follows:

\begin{eqnarray}
q_{-}(t) &\rightarrow q_{-}(t) - \bar{q}_{-}(t) = \delta q{-}(t)\nonumber\\
p_{-}(t) &\rightarrow p_{-}(t) - \bar{p}_{-}(t) = \delta p{-}(t)\;.
\end{eqnarray}

The generalized synchronization measure in the quantum regime can then be expressed only in terms of these fluctuations, as 
\begin{equation}
S_{q} \equiv \left\langle \delta q_{-}^{2}(t) + \delta p_{-}^{2}(t) \right\rangle^{-1} \label{eq:sq}
\end{equation}
In case of a constant phase between the limit cycles, we employ the so-called ``quantum $\phi$ synchronization" \cite{qiao2020quantum,PhysRevA.109.023502}, defined as 
\begin{equation}
S_{q}^{\phi}(t) \equiv \left\langle \delta q_{-}^{\phi}(t)^{2} + \delta p_{-}^{\phi}(t)^{2} \right\rangle^{-1} \label{eq:sqm}\;,
\end{equation}

The $\phi$ error operators are defined as $q_{-}^\phi(t)=\frac{1}{\sqrt{2}}\left[q_1^\phi(t)-q_2^\phi(t)\right]$ and $p_{-}^\phi(t)=\frac{1}{\sqrt{2}}\left[p_1^\phi(t)-p_2^\phi(t)\right]$with  \\
\begin{eqnarray}
q_j^\phi(t) & =q_j(t) \cos \left(\phi_j\right)+p_j(t) \sin \left(\phi_j\right),\nonumber\\
p_j^\phi(t) & =p_j(t) \cos \left(\phi_j\right)-q_j(t) \sin \left(\phi_j\right).
\end{eqnarray}
where, $\phi = \phi_{2} - \phi_{1}$ represents the phase difference between the two limit cycles, with $\phi_{j} = \tan^{-1}\left[\frac{\Bar{p}_{j}(t)}{\Bar{q}_{j}(t)}\right]$.

The initial states of the oscillators are ideally approximated as Gaussian distributions centered at their respective mean positions, with minimal positional uncertainties. The fluctuation dynamics of the system are governed by a set of linearized equations, ensuring that the Gaussian characteristics are maintained during the evolution. The Gaussian states can be fully characterized by their covariance matrices. We exploit this property to compute correlations between quantum fluctuations of the quadratures and to evaluate relevant synchronization measures. The matrix $C$ of the covariance matrices follows the linear differential equation, as given by
\begin{small}
\begin{equation}
\dot{C}(t)=M(t)C(t) + C(t) M(t)^{T}+D\;,
\end{equation}
\end{small}
where the elements of $C$ can be identified as $C_{ij}= \left[\left\langle Y_i(t)Y_j(t)+Y_j(t)Y_i(t)\right\rangle \right]/2 $ and the diffusion matrix $D$ is given by

\begin{small}
\begin{equation}
\begin{aligned}
D =& {\rm diag}\Bigg[ V_1, V_1, V_2, V_2 \Bigg]
\label{dmat}
\end{aligned}
\end{equation}
\end{small}
where $V_j=\left([{\rm Re}(U_j)]^2+[{\rm Im}(U_j)]^2\right)\left( \bar{n}_{m} + 0.5 \right)$, and  $j\in 1,2$.

In the matrix $C$, every diagonal element represents the $2 \times 2$ covariance matrix for the respective mode and every non-diagonal element $C_{ij}$ represents the $2 \times 2$ matrix of inter-mode covariance.

The complete quantum synchronization $S_{q}^\phi(t)$ can then be expressed in a concise form as

\begin{small}
\begin{equation}
\begin{aligned}
S_{q}^\phi(t) &= 2 \left[C_{11}(t) + C_{22}(t) + C_{33}(t) + C_{44}(t) + 2 \sin{\phi}C_{23}(t) \right. \\
& \quad \left. - 2\sin{\phi}C_{14}(t) - 2\cos{\phi}C_{13}(t) - 2\cos{\phi}C_{24}(t)\right]^{-1}
\end{aligned}
\end{equation}
\end{small}

\section{Numerical Results} 

In this section, we will discuss the degree of quantum synchronization between the coupled spin chains by numerically solving the Eqs. (\ref{beta1eqn}), (\ref{beta2eqn}) and (\ref{c1}). To analyze the classical synchronization between the spin-chain oscillators, we numerically solve the Eqs. (\ref{beta1eqn}) and (\ref{beta2eqn}) for $\beta_j(t)$. We then introduce the position and linear momentum quadratures of the two oscillators: $q_j=\frac{1}{\sqrt{2}}\left(b_j^{\dagger}+b_j\right)$, $p_j=\frac{\iota}{\sqrt{2}}\left(b_j^{\dagger}-b_j\right)$. We show the temporal variation of these quadratures in a phase space diagram in Figs. \ref{limitcycle}. We can see from Fig. \ref{pqa} that for $\lambda = 0$ (i.e., when the spins in the chains do not interact directly, but via the central spin) and an equal number of spins (i.e., $N_{1}=N_{2}$), the spin-chain oscillators exhibit the same limit cycle at the long times $t \approx 10^4\tau$. We assumed in this diagram that these oscillators are initialized from $q_j(0)=p_j(0) = 0$. The mean values of quadratures also become equal, i.e., $\bar{q}_1(t) = \bar{q}_2(t)$ and $\bar{p}_1(t) = \bar{p}_2(t)$, as shown in Figs. \ref{q1q2a} and \ref{p1p2a}. This refers to a classical synchronization between two finite-size many-spin chains.

Next, we analyze the case when the sizes of the spin chains are unequal and $\lambda = 0$. Figure \ref{pqb} shows the time evolution of limit-cycle trajectories in phase space, and we can easily see that this evolution is different from Figure \ref{pqa}. As shown in Figures \ref{q1q2b} and \ref{p1p2b}, at the long times, the mean values of position and momenta are in the same phase, but their amplitudes are different from each other. This indicates classical synchronization between the spin chains. 

Next, we consider the case when the spin sizes are equal and the coupling between the spins in the chains is nonzero, i.e., $\lambda \neq 0$). The figure \ref{pqc} shows the time evolution of limit-cycle trajectories in phase space. Even though the mean positions $\bar{q}_{1}$ and $\bar{q}_{2}$ exhibit steady oscillations, their evolutions are not identical, as shown in Figure \ref{q1q2c}. A similar trend can be found in the evolutions of the mean momenta $\bar{p}_{1}$ and $\bar{p}_{2}$ as well [see Figure \ref{p1p2c}]. Their mean values have a small phase difference, $\phi = 1.049$. In Fig. \ref{Sq}, we have plotted the measure of quantum complete synchronization between the coupled spin-chains, when $N_1=N_2$ and $\lambda=0.2$. It can be found from this figure, that the optimal value of the quantum synchronization marker is unity (i.e., $S_{q}=1$) in the steady state, which refers to complete synchronization between the chains. 

We have observed that the optimal value of $S_{q}$ remains unity for all the above cases
[see Fig. \ref{limitcycle}]. This establishes the fact that the synchronization remains robust against the variation of the number of spins, and also in the presence of inter-spin coupling. To check the robustness of synchronization, we plotted in Fig. \ref{synchronization temp}  the time-averaged value of quantum synchronization, $\bar{S}_{q}$ with respect to the mean number of thermal phonons $\bar{n}_{m}$ for $N_{1}=N_{2}=5$ and $\lambda=0.2$. We find that as $\bar{n}_{m}$ increases, the synchronization deteriorates.

Our investigation uncovers a compelling correspondence between the chains' distinctive limit cycle trajectories and the emergence of quantum synchronization. The boson-number-dependent nonlinear interaction between the chains is crucial in building this synchronization, in the presence of linear as well as nonlinear damping, unlike in the case of van der Pol oscillators, in which linear coupling is associated with nonlinear damping. Such a nonlinear Hamiltonian is expected to build up non-Gaussian entangled states of the oscillators. To correlate this entanglement with quantum synchronization, one may invoke the higher-order entanglement criteria, as formulated by Shchukin and Vogel in their seminal work \cite{shchukin2005inseparability}. However, in such a case, the covariance analysis based on the Gaussian approximation needs modification. Our results are valid at the Gaussian limit. We have also verified that the two oscillators, in our case, are not entangled, yet completely synchronized.

\section{CONCLUSION AND REMARKS} 

We have established a comprehensive theoretical framework to rigorously investigate the phenomena of quantum synchronization between two finite-size spin chains. These chains indirectly interact with each other, via their common coupling with a central spin chain. The effective coupling between the two chains becomes nonlinear. Further, their dynamics are also affected by both linear and non-linear dissipation. 

While spin synchronization is often studied in terms of Q function or spin-spin correlation, we have put forward a new approach in this regard. We replace the collective spin with an equivalent pseudo-bosonic operator, using the Holstein-Primakoff transformation. Utilizing the synchronization measure, developed for bosonic system, we have quantitatively assessed the degree of synchronization between the two coupled spin chains. 

Our numerical simulations demonstrate that, under optimal system parameters, the coupled spin chains can attain both classical and perfect quantum synchronization. Notably, perfect quantum synchronization is indicated by the synchronization measure $S_{q}$ achieving a value of 1, signifying complete synchronization at the quantum level.
We further show that the quantum synchronization is robust against the variation of the number of spins in the chain and the inter-spin coupling. However, the thermal noise deteriorates the quantum synchronization. 

Our work significantly advances the understanding of synchronization dynamics in multi-spin systems. Further, we have put forward a generalized measure of synchronization, that is valid for both bosons and fermions. This work is expected to provide a new perspective of the collective oscillatory dynamics of many-spin systems.

\section{ACKNOWLEDGMENTS}

One of us (J.G.) acknowledges the financial support provided by the Department of Science and Technology-Innovation in Science Pursuit for Inspired Research (DST-INSPIRE) during this work.

\appendix

\section{Langevin's equations}

\begin{equation}\label{b1eqn}
\begin{small}
\begin{aligned}
\dot{b}_1 &= \iota \theta{b}_1 - \frac{\gamma_l}{2} {b}_1 + {\left[\sqrt{\gamma_l}+\frac{1}{N_1} \sqrt{\gamma_{nl}}\left(2 b_1^{\dagger} b_1-b_1^2\right)\right] b_{\text {in}}^{(1)}} \\
& -\frac{\gamma_{nl}}{N_1^2} b_1^{\dagger2} b_1^3-\frac{2\gamma_{nl}}{N_1^2} b_1^{\dagger} b_1^2-\frac{2}{N_1} \sqrt{\gamma_l\gamma_{nl}} b_1^{\dagger} b_1^2+\frac{\iota\left\langle\sigma_z\right\rangle }{X} \\
 &\quad \times \left\{g_1^2\left(\frac{b_1^{\dagger} b_1}{2N_1}-1\right) \left(1-\frac{b_1^{\dagger} b_1}{4N_1}\right) b_1 + g_1^2 b_1^{\dagger} \left(1-\frac{b_1^{\dagger} b_1}{4N_1}\right)\frac{b_1^2}{4N_1}\right. \\
&\quad +g_1 g_2 b_2^{\dagger} \exp\left[-R_2 \left(b_2^{\dagger} b_2 + b_2 b_2^{\dagger}\right)\right] \left(1 - \frac{b_2^{\dagger} b_2}{4N_2}\right) \\
&\quad \times \left[\frac{b_1}{4 N_1} \exp\left[R_1 \left(b_1^{\dagger} b_1 + b_1 b_1^{\dagger}\right)\right] b_1 \right. \\
&\quad - 2R_1 \left(1 - \frac{b_1^{\dagger} b_1}{4 N_1}\right) \exp\left[R_1 \left(b_1^{\dagger} b_1 + b_1 b_1^{\dagger}\right)\right] b_1^2 \Bigg]\\
&\quad+ g_1 g_2 \left[ b_1^{\dagger} \exp\left[-R_1 \left(b_1^{\dagger} b_1 + b_1 b_1^{\dagger}\right)\right] \frac{b_1}{4 N_1} \right. \\
&\quad +2R_1 b_1^{\dagger} \exp\left[-R_1 \left(b_1^{\dagger} b_1 + b_1 b_1^{\dagger}\right)\right] b_1 \left(1 - \frac{b_1^{\dagger} b_1}{4 N_1}\right) \\
&\quad - \exp\left[-R_1 \left(b_1^{\dagger} b_1 + b_1 b_1^{\dagger}\right)\right] \left(1 - \frac{b_1^{\dagger} b_1}{4 N_1}\right) \Bigg] \\
&\quad \left. \times \left(1 - \frac{b_2^{\dagger} b_2}{4 N_2}\right) \exp\left[R_2 \left(b_2^{\dagger} b_2 + b_2 b_2^{\dagger}\right)\right] b_2\right\}
\end{aligned}
\end{small}
\end{equation}

\begin{equation}\label{b2eqn}
\begin{small}
\begin{aligned}
& \dot{b}_2=-\iota \theta{b}_2-\frac{\gamma_l}{2}{b}_2+ {\left[\sqrt{\gamma_l}+\frac{1}{N_2} \sqrt{\gamma_{nl}}\left(2 b_2^{\dagger} b_2-b_2^2\right)\right] b_{\text {in}}^{(2)}} \\
& -\frac{\gamma_{nl}}{N_2^2} b_2^{\dagger2} b_2^3-\frac{2\gamma_{nl}}{N_2^2} b_2^{\dagger} b_2^2-\frac{2}{N_2} \sqrt{\gamma_l\gamma_{nl}} b_2^{\dagger}+\frac{\iota\left\langle\sigma_z\right\rangle}{X} \\
&\quad \times \left\{g_2^2\left(\frac{b_2^{\dagger} b_2}{2N_2}-1\right) \left(1-\frac{b_2^{\dagger} b_2}{4N_2}\right) b_1 + g_2^2 b_2^{\dagger} \left(1-\frac{b_2^{\dagger} b_2}{4N_2}\right)\frac{b_2^2}{4N_2}\right. \\
&\quad +g_1 g_2 b_1^{\dagger} \exp\left[-R_1 \left(b_1^{\dagger} b_1 + b_1 b_1^{\dagger}\right)\right] \left(1 - \frac{b_1^{\dagger} b_1}{4N_1}\right) \\
&\quad \times \left[\frac{b_2}{4 N_2} \exp\left[R_2 \left(b_2^{\dagger} b_2 + b_2 b_2^{\dagger}\right)\right] b_2 \right. \\
&\quad - 2R_2 \left(1 - \frac{b_2^{\dagger} b_2}{4 N_2}\right) \exp\left[R_2\left(b_2^{\dagger} b_2 + b_2 b_2^{\dagger}\right)\right] b_2^2 \Bigg]\\
&\quad+ g_1 g_2 \left[ b_2^{\dagger} \exp\left[-R_2 \left(b_2^{\dagger} b_2 + b_2 b_2^{\dagger}\right)\right] \frac{b_2}{4 N_2} \right. \\
&\quad +2R_2 b_2^{\dagger} \exp\left[-R_2 \left(b_2^{\dagger} b_2 + b_2 b_2^{\dagger}\right)\right] b_2 \left(1 - \frac{b_2^{\dagger} b_2}{4 N_2}\right) \\
&\quad - \exp\left[-R_2 \left(b_2^{\dagger} b_2 + b_2 b_2^{\dagger}\right)\right] \left(1 - \frac{b_2^{\dagger} b_2}{4 N_2}\right) \Bigg] \\
&\quad \left. \times \left(1 - \frac{b_1^{\dagger} b_1}{4 N_1}\right) \exp\left[R_1 \left(b_1^{\dagger} b_1 + b_1 b_1^{\dagger}\right)\right] b_1\right\}
\end{aligned}
\end{small}
\end{equation}

\section{Equations in mean field approximations}

\begin{small}
\begin{equation} \label{beta1eqn}
\begin{aligned}
\dot{\beta}_1 &= \iota\theta\beta_1 - \frac{\gamma_l}{2}\beta_1 - \frac{\gamma_{nl}}{N_1^2}|\beta_1|^4\beta_1 - 2\frac{\gamma_{nl}}{N_1^2}|\beta_1|^2\beta_1 \\
&\quad - \frac{2}{N_1}\sqrt{\gamma_l\gamma_{nl}}|\beta_1|^2\beta_1 + \frac{\iota\left\langle\sigma_z\right\rangle}{X} \Bigg[g_1^2|\beta_1|^2\frac{\beta_1}{N_1} - \frac{3g_1^2|\beta_1|^4\beta_1}{16N_1^2} \\
&\quad\quad - g_1^2\beta_1 + g_1g_2\beta_2^* e^{-R_2a_2 + R_1a_1} A_2 \left( \frac{\beta_1^2}{4N_1} - 2R_1A_1\beta_1^2 \right) \\
&\quad\quad + g_1g_2 A_2 e^{R_2a_2 -R_1a_1} \beta_2 \left( \frac{|\beta_1|^2\beta_1}{4N_1} + 2R_1A_1|\beta_1|^2 - A_1 \right) \Bigg]
\end{aligned}
\end{equation}

\begin{equation} \label{beta2eqn}
\begin{aligned}
\dot{\beta}_2 &= -\iota\theta\beta_2 - \frac{\gamma_l}{2}\beta_2 - \frac{\gamma_{nl}}{N_2^2}|\beta_2|^4\beta_2 - 2\frac{\gamma_{nl}}{N_2^2}|\beta_2|^2\beta_2\\
&\quad - \frac{2}{N_2}\sqrt{\gamma_l\gamma_{nl}}|\beta_2|^2\beta_2 + \frac{\iota\left\langle\sigma_z\right\rangle}{X} \Bigg[g_2^2|\beta_2|^2\frac{\beta_2}{N_2} - \frac{3g_2^2|\beta_2|^4\beta_2}{16N_2^2}\\
&\quad\quad - g_2^2\beta_2  + g_1g_2\beta_1^* e^{-R_1a_1 + R_2a_2} A_1 \left( \frac{\beta_2^2}{4N_2} - 2R_2A_2\beta_2^2 \right) \\
&\quad\quad + g_1g_2 A_1 e^{R_1a_1 -R_2a_2} \beta_1 \left( \frac{|\beta_2|^2\beta_2}{4N_2} + 2R_2A_2|\beta_2|^2 - A_2 \right) \Bigg]
\end{aligned}
\end{equation}
\end{small}
where
$A_j = 1 - \frac{|\beta_j|^2}{4N_j}$ and $a_j = 1+ 2|\beta_j|^2$.

\section{Fluctuation Equations}
\begin{small}
\begin{eqnarray} \label{c1}
\delta\dot{b}_1 &=& E_1\delta{b}_1 + E_2\delta{b}_1^{\dagger} + E_3\delta{b}_2 + E_4\delta{b}_2^{\dagger}+U_1 b_{\text{in}}^{(1)}+F_1\;,\nonumber\\
\delta\dot{b}_2 &=& E_5\delta{b}_1 + E_6\delta{b}_1^{\dagger} + E_7\delta{b}_2 + E_8\delta{b}_2^{\dagger}+U_2 b_{\text{in}}^{(2)}+F_2\;.
\end{eqnarray}
\end{small}
where
\[
 \quad \delta{b}_j = \frac{\delta{q}_j + i\delta{p}_j}{\sqrt{2}}, \quad j = [1,2]
\]
and 

{\footnotesize

\begin{align*}
E_1 = & \iota\theta 
      - \frac{\gamma_l}{2} 
      - \frac{3}{N_1} \gamma_{nl} |\beta_1|^4 
      - \frac{4}{N_1^2} \gamma_{nl} |\beta_1|^2 
      - \frac{4}{N_1} \sqrt{\gamma_l \gamma_{nl}} |\beta_1|^2 \\
    & + \frac{\iota\left\langle\sigma_z\right\rangle}{X} \Bigg[ 
        g_1 \left( \frac{3|\beta_1|^2}{2N_1} - \frac{9|\beta_1|^4}{16N_1^2} - 1 \right) \\
    & + g_1 g_2 A_2 \beta_1\beta_2^* e^{-R_2 a_2} \left\{
        \left( \frac{1}{2N_1} \right) e^{R_1 a_1} + R_1 \left( \frac{\beta_1^*}{2N_1} \right) \right. \\
    & +\left. \frac{R_1 |\beta_1|^2 e^{R_1 a_1}}{2N_1}
      - 4 R_1^2 A_1 |\beta_1|^2
      - 4 R_1 A_1  e^{R_1 a_1}\right\}  \\
    & + g_1 g_2 A_2 \beta_1^* \beta_2e^{R_2 a_2} \left\{
        -R_1 \left( \frac{|\beta_1|^2}{2N_1} \right)
        + \frac{1}{N_1} e^{-R_1 a_1} \right. \\
    & - 4 R_1^2 |\beta_1|^2 A_1 
      + 2 R_1 e^{-R_1 a_1} A_1 
    + 2 R_1 A_1  \\
    &   \left.- \frac{R_1 e^{-R_1 a_1} |\beta_1|^2 }{2N_1}
       \right\} \Bigg]\;,
\end{align*}
}

{\footnotesize
\begin{align*}
E_2 = & - \frac{2}{N_1^2} \gamma_{nl} |\beta_1|^2 \beta_1^2 - \frac{2}{N_1^2} \gamma_{nl} \beta_1^2 - \frac{2}{N_1} \sqrt{\gamma_l \gamma_{nl}} \beta_1^2\\
& + \frac{\iota\left\langle\sigma_z\right\rangle}{X} \Bigg[ 
        g_1 \left(\frac{3|\beta_1|^2}{2N_1} - \frac{9|\beta_1|^4}{16N_1^2} - 1 \right) \\
      & + g_1 g_2 R_1A_2\beta_1^2\beta_2^* e^{-R_2 a_2} \left\{
        \left(\frac{1}{2N_1}\right) + \left( \frac{\beta_1}{2N_1}\right)e^{R_1 a_1} -4 R_1 A_1 \beta_1
        \right\} \\
      & + g_1 g_2 A_2\beta_1\beta_2 e^{R_2 a_2} \left\{
        \frac{1}{2N_1}e^{-R_1 a_1} - R_1 \left( \frac{|\beta_1|^2 }{2N_1}\right) + 2R_1 A_1 e^{-R_1 a_1} \right. \\
      & - 4 A_1 R_1^2 |\beta_1|^2 - \left.R_1 \left( \frac{|\beta_1|^2}{2N_1}\right)e^{-R_1 a_1} + 2R_1  A_1 
        \right\}
      \Bigg]\;,
\end{align*}
}

{\footnotesize
\begin{align*}
E_3 = &
\frac{\iota g_1g_2\left\langle\sigma_{\mathrm{z}}\right\rangle}{X}\Bigg[\beta_1^2\beta_2^{*2}e^{R_1 a_1}\left(2R_2 A_2+\frac{e^{-R_2 a_2}}{4N_2}\right)\left(2 R_1 A_1-\frac{1}{4N_1}\right)\\
& +e^{-R_1 a_1}\left(2A_2R_2\left|\beta_2\right|^2-\frac{ e^{R_2 a_2}\left|\beta_2\right|^2}{4N_2}+A_2e^{R_2a_2}\right)\\
&\times\left(\frac{\left|\beta_1\right|^2}{4N_1}+ 2 R_1A_1 \left|\beta_1\right|^2 -A_1\right)\Bigg]\;,
\end{align*}

{\footnotesize
\begin{align*}
E_4 = & \frac{\iota
g_1g_2\left\langle\sigma_{\mathrm{z}}\right\rangle}{X}\Bigg[e^{R_1 a_1} \beta_1^2\left(-2 R_2A_2\left|\beta_2\right|^2 +A_2e^{-R_2 a_2}-\frac{e^{-R_2 a_2}\left|\beta_2\right|^2}{4N_2}\right)\\
&\left(\frac{1}{4N_1}-2 R_1 A_1\right)+  e^{-R_1 a_1}\beta_2^2\left(2R_2A_2 -\frac{e^{R_2 a_2}}{4N_2}\right)\\
&\left(\frac{\left|\beta_1\right|^2}{4N_1}+2 R_1A_1  \left|\beta_1\right|^2-A_1 \right)\Bigg]\;,
\end{align*}
}

{\footnotesize
\begin{align*}
E_5 = & \frac{\iota g_1g_2\left\langle\sigma_{\mathrm{z}}\right\rangle}{X} \Bigg[e^{R_2a_2}\beta_2^2\beta_1^{*2} \left(2R_1 A_1 +\frac{e^{-R_1a_1}}{4N_1} \right) \left(2 R_2 A_2-\frac{1}{4 N_2} \right) \\
& +e^{-R_2a_2} \left( 2 A_1 R_1 \left|\beta_1\right|^2 - \frac{e^{R_1a_1} \left|\beta_1\right|^2}{4N_1} + A_1 e^{R_1a_1} \right) \\
& \left( \frac{\left|\beta_2\right|^2}{4N_2} + 2 A_2 R_2 \left|\beta_2\right|^2 - A_2  \right) \Bigg]\;,
\end{align*}
}

{\footnotesize
\begin{align*}
E_6 = & \frac{\iota g_1g_2\left\langle\sigma_{\mathrm{z}}\right\rangle}{X} \Bigg[e^{R_2a_2} \beta_2^2 \left(-2R_1A_1 \left|\beta_1\right|^2 + A_1 e^{-R_1a_1} - \frac{e^{-R_1a_1} \left|\beta_1\right|^2}{4N_1} \right) \\
&\left( \frac{1}{4N_2} - 2 R_2 A_2 \right)+ e^{-R_2a_2}\beta_1^2\left( 2R_1A_1 - \frac{ e^{R_1a_1}}{4N_1} \right)\\
&\left( \frac{ \left|\beta_2\right|^2}{4N_2} + 2R_2A_2   \left|\beta_2\right|^2 - A_2  \right) \Bigg]\;,
\end{align*}
}

{\footnotesize
\begin{align*}
E_7 = & -\iota\theta 
      - \frac{\gamma_l}{2} 
      - \frac{3}{N_2} \gamma_{nl} |\beta_2|^4 
      - \frac{4}{N_2^2} \gamma_{nl} |\beta_2|^2 
      - \frac{4}{N_2} \sqrt{\gamma_l \gamma_{nl}} |\beta_2|^2 \\
    & + \frac{\iota\left\langle\sigma_z\right\rangle}{X} \Bigg[ 
        g_2 \left( \frac{3|\beta_2|^2}{2N_2} - \frac{9|\beta_2|^4}{16N_2^2} - 1 \right) \\
    & + g_1 g_2 A_1 \beta_1^* \beta_2e^{-R_1 a_1} \left\{
        \left( \frac{1}{2N_2} \right) e^{R_2 a_2} + R_2 \left( \frac{\beta_2^*}{2N_2} \right)
    \right. \\
    & +\left. \frac{R_2 |\beta_2|^2 e^{R_2 a_2}}{2N_2}
      - 4 R_2^2 A_2 |\beta_2|^2  
      - 4 R_2 A_2 e^{R_2 a_2}  \right\} \\
    & + g_1 g_2 A_1 \beta_1\beta_2^* e^{R_1 a_1} \left\{
        -R_2 \left( \frac{|\beta_2|^2}{2N_2} \right)  
        + \frac{1}{2N_2} e^{-R_2 a_2} \right. \\
    & - 4 R_2^2 |\beta_2|^2 A_2 
      + 2 R_2 e^{-R_2 a_2} A_2 
     + 2 R_2 A_2 \\
    &  \left. - \frac{R_2 e^{-R_2 a_2} |\beta_2|^2 }{2N_2}
       \right\} \Bigg]\;,
\end{align*}
}

{\footnotesize
\begin{align*}
E_8 = & - \frac{2}{N_2^2} \gamma_{nl} |\beta_2|^2 \beta_2^2 - \frac{2}{N_2^2} \gamma_{nl} \beta_2^2 - \frac{2}{N_2} \sqrt{\gamma_l \gamma_{nl}} \beta_2^2 \\
& + \frac{\iota\left\langle\sigma_z\right\rangle}{X} \Bigg[ 
        g_2 \left(\frac{3|\beta_2|^2}{2N_2} - \frac{9|\beta_2|^4}{16N_2^2} - 1 \right) \\
& + g_1 g_2R_2A_1\beta_1^*\beta_2^2 e^{-R_1 a_1} \left\{\left(\frac{1}{2N_2}\right) + \left( \frac{\beta_2}{2N_2}\right)e^{R_2 a_2} -4 R_2A_2 \beta_2\right\}\\
& + g_1 g_2 A_1\beta_1 \beta_2 e^{R_1 a_1}  \left\{\frac{1}{2N_2}e^{-R_2 a_2}- R_2 \left( \frac{|\beta_2|^2 }{2N_2}\right) +2R_2A_2 e^{-R_2 a_2}\right. \\
& - 4 A_2 R_2^2|\beta_2|^2  - \left.R_2 \left( \frac{|\beta_2|^2 }{2N_2}\right)e^{-R_2 a_2} + 2R_2  A_2 \right\}
      \Bigg]\;.
\end{align*}
}

{\footnotesize
\begin{align*}
F_1 &= \frac{i g_1\left\langle\sigma_z\right\rangle}{X} \Bigg\{ 
    -2 A_1 \beta_1 
    + \frac{3 \beta_1 \left|\beta_1\right|^2}{2N_1} 
    - 2 \beta_1 \\
    & \quad + \beta_1^2 \beta_2^* e^{R_1 a_1 - R_2 a_2} \Bigg[
        \frac{3 A_2 + 1}{2N_1} 
        - 4 R_1 \left( A_1 + A_2 \right) 
        - 10 R_1 A_1 A_2 
    \Bigg]\\
    & \quad+e^{R_2 a_2 - R_1 a_1}\Bigg[\left(\frac{5|\beta_1|^2}{2N_1}+20|\beta_1|^2R_1A_1+4|\beta_1|^2R_1-12A_1-1\right)\\
    &+2\beta_2\left(\frac{|\beta_1|^2}{4N_1}+4|\beta_1|^2R_1A_1-A_1\right)\Bigg]
\Bigg\}.
\end{align*}
}

{\footnotesize
\begin{align*}
F_2 &= \frac{i g_1\left\langle\sigma_z\right\rangle}{X} \Bigg\{ 
    -2 A_2 \beta_2 
    + \frac{3 \beta_2 \left|\beta_2\right|^2}{2N_2} 
    - 2 \beta_2 \\
    & \quad + \beta_2^2 \beta_1^* e^{R_2 a_2 - R_1 a_1} \Bigg[
        \frac{3 A_1 + 1}{2N_2} 
        - 4 R_2 \left( A_1 + A_2 \right) 
        - 10 R_2 A_1 A_2 
    \Bigg]\\
    & \quad+e^{R_1 a_1 - R_2 a_2}\Bigg[\left(\frac{5|\beta_2|^2}{2N_2}+20|\beta_2|^2R_2A_2+4|\beta_2|^2R_2-12A_2-1\right)\\
    &+2\beta_1\left(\frac{|\beta_2|^2}{4N_2}+4|\beta_2|^2R_2A_2-A_2\right)\Bigg]
\Bigg\}.
\end{align*}
}

\bibliographystyle{ieeetr}

\bibliography{main}

\end{document}